\documentclass{article}
\usepackage{spconf,amsmath,amssymb,graphicx,url,times,color,booktabs,multirow}
\newcommand\linesubsec[1]{\vspace{0.8mm}\noindent\textbf{#1 --- }}

\usepackage[dvipsnames]{xcolor}

\title{Direct design of biquad filter cascades with deep \\ learning by sampling random polynomials}
\name{Joseph T. Colonel$^{1*}$ \quad
      Christian J. Steinmetz$^{1*}$ \quad
      Marcus Michelen$^{2}$  \quad
      Joshua D. Reiss$^{1}$}
\address{$^1$Centre for Digital Music,  Queen Mary University of London\\ 
         $^2$Department of Mathematics, Statistics, and Computer Science, University of Illinois at Chicago\\
}

\begin{document}
\ninept
\maketitle

\def\thefootnote{*}\footnotetext{ \vspace{-0.0cm}These authors contributed equally to this work.}\def\thefootnote{\arabic{footnote}}

\begin{abstract}
Designing infinite impulse response filters to match an arbitrary magnitude response requires specialized techniques.
Methods like modified Yule-Walker are relatively efficient, but may not be sufficiently accurate in matching high order responses.
On the other hand, iterative optimization techniques often enable superior performance, but come at the cost of longer run-times and are sensitive to initial conditions, requiring manual tuning.
In this work, we address some of these limitations by learning a direct mapping from the target magnitude response to the filter coefficient space with a neural network trained on millions of random filters.
We demonstrate our approach enables both fast and accurate estimation of filter coefficients given a desired response. 
We investigate training with different families of random filters, and find training with a variety of filter families enables better generalization when estimating real-world filters, using head-related transfer functions and guitar cabinets as case studies. 
We compare our method against existing methods including modified Yule-Walker and gradient descent and show our approach is, on average, both faster and more accurate.

\end{abstract}
\section{Introduction}
\label{sec:intro}

Infinite impulse response (IIR) filters have a variety of applications, such as control systems, time series forecasting, and audio signal processing~\cite{dorf2011modern,hamilton1994time,valimaki2016all}. 
In audio applications, digital IIR filters are often used for equalisation, including tone matching, feedback reduction, and room compensation~\cite{ramos2006filter}.
Classical methods for designing digital IIR filters are generally restricted to specific prototypes, e.g. designing a lowpass filter with minimum passband ripple~\cite{selesnick1998generalized}. 
However, some applications require designing a filter that achieves an arbitrary magnitude and/or phase response.
Classical methods for this task include the modified Yule-Walker (MYW) estimation~\cite{chan1982spectral}, least squares approaches~\cite{kobayashi1990design, lang1998weighted},
linear programming~\cite{rabiner1974linear}, Steiglitz-McBride~\cite{stoica1981steiglitz}, and gradient-based optimization methods~\cite{dodds2020a}. 

However, these approaches have drawbacks that may limit their application in scenarios that require high accuracy, fast estimation, or both. 
For example, while MYW can be performed quickly with a small number of operations, it may produce inaccurate results for more challenging target responses.
On the other hand, iterative methods often provide greater accuracy and can be tailored with customized loss functions.
Although, this comes with higher computational cost due to need for multiple gradient update operations.
In addition, since this optimization process is generally non-convex, performance is often very sensitive to initial conditions and also may suffer from getting stuck in local minima~\cite{dodds2020a, nercessian2020neural}. 

\newpage
Recently, there has been interest in integrating deep learning approaches for filter design.
The parallels between recurrent neural networks (RNNs) and IIR filters have been exploited to learn arbitrary filters from data~\cite{kuznetsov2020differentiable, pepe2020designing, ramirez2018end}. 
While these networks simulate the sample-by-sample operations of a digital IIR filter, they can be slow and difficult to train due to their recursive nature, which requires many gradient steps through time.
Other approaches have instead been trained to directly estimate the parameters of graphic~\cite{8809819} and parametric equalizers~\cite{nercessian2020neural, 9415855} given a desired magnitude response.
While these approaches avoid the need for iterative estimation, they are potentially restricted by the IIR filter prototypes they estimate.
\begin{figure}[]
    \centering
    \vspace{-0.25cm}
    \includegraphics[width=0.99\linewidth,trim={0.6cm 0.0cm 0.0cm 0.0cm},clip]{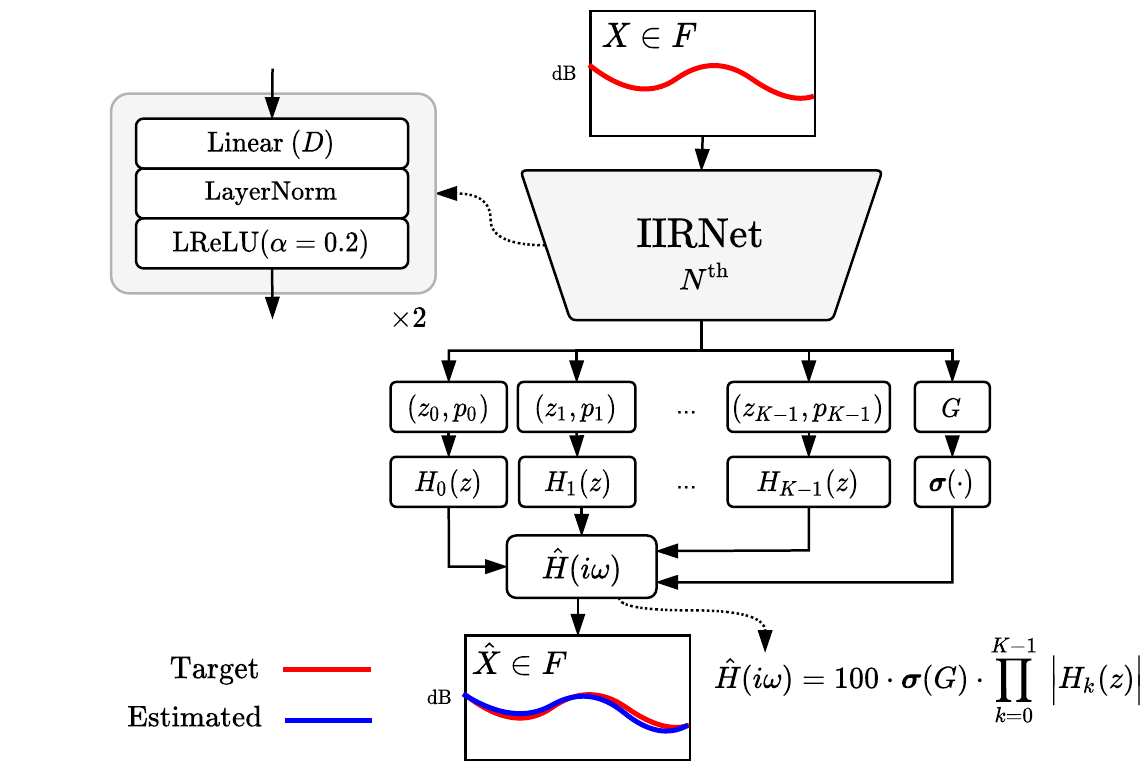}
    \vspace{-0.2cm}
\caption{IIRNet predicts $N^\textrm{th}$ order filters composed of $K$ biquads.}
    \label{fig:iirnet-arch}
    \vspace{-0.52cm}
\end{figure}

We aim to address these limitations by constructing a model capable of learning the mapping from a desired arbitrary magnitude response directly to the coefficients of an IIR filter, removing the need for iterative optimization.
To achieve this, we propose IIRNet, a neural network trained with randomly generated filters to estimate a cascade of biquads given a desired magnitude response, as shown in Figure~\ref{fig:iirnet-arch}.
We investigate methods for generating a diverse random filters using knowledge of the behavior of random polynomials. 

The contributions of this work are as follows. 
First, we build on previous work with an improved domain-inspired architecture for stable training of neural network biquad filter cascade estimation to match arbitrary magnitude responses named IIRNet. 
Second, we propose a training regime that involves generating random filters from a diverse range of families of random polynomials and empirically demonstrate how training with this set of diverse polynomial families enables better generalization in real-world tasks. 
Finally, we demonstrate that our model combined with a training regime covering a range of random filter families generalizes to real-world filter estimation, and is shown to, on average, outperform MYW and gradient-based techniques in terms of run-time and accuracy. 

\begin{figure*}[ht] 
    \begin{minipage}{.16\textwidth}
        \centering
        \includegraphics[width=\linewidth,trim={1.7cm 0.3cm 2.6cm 0.3cm},clip]{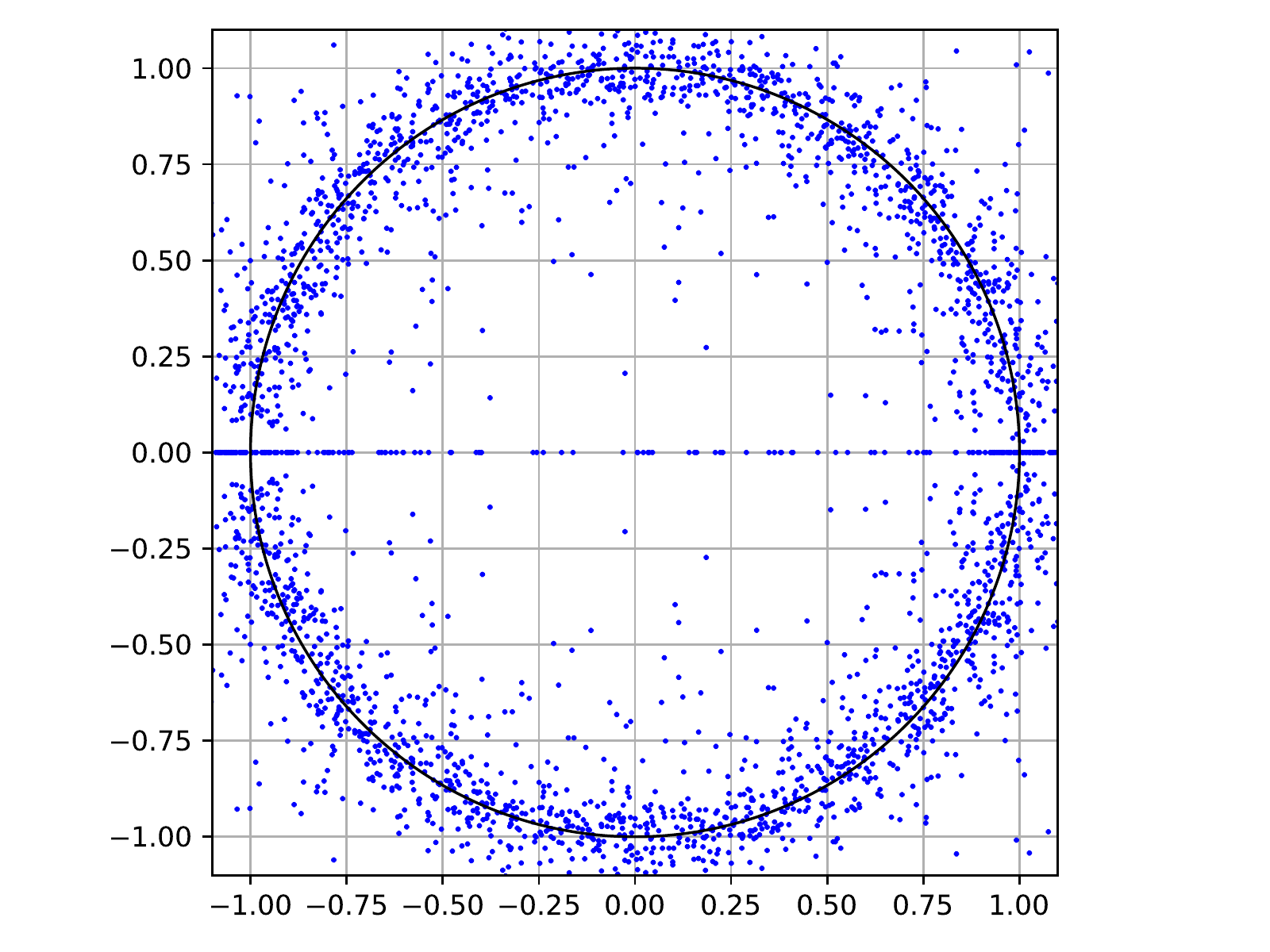} \\ (A)
    \end{minipage}
    \begin{minipage}{.16\linewidth}
        \centering
        \includegraphics[width=\linewidth,trim={1.7cm 0.3cm 2.6cm 0.3cm},clip]{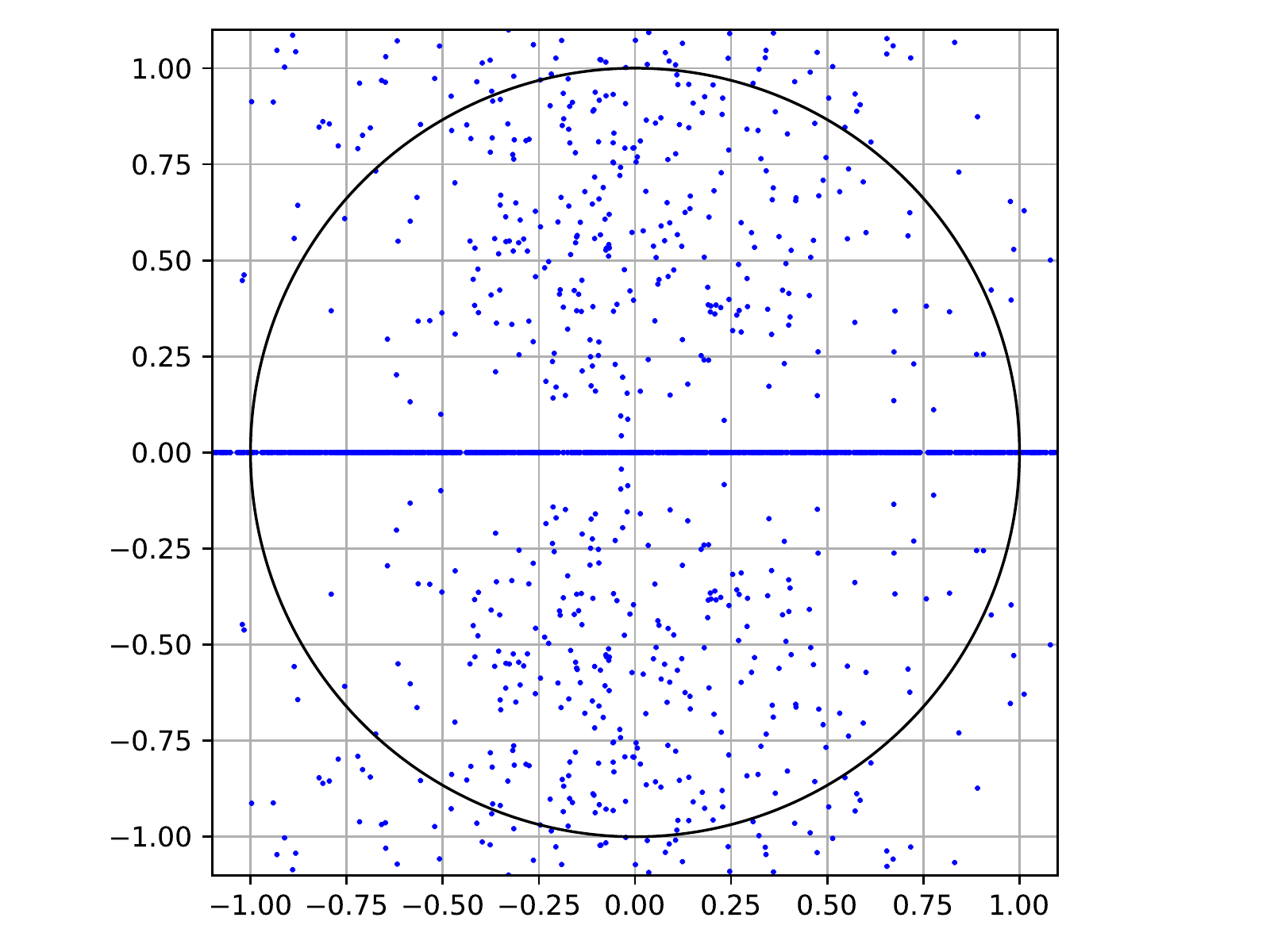} \\ (B)
    \end{minipage}
    \begin{minipage}{.16\linewidth}
        \centering
        \includegraphics[width=\linewidth,trim={1.7cm 0.3cm 2.6cm 0.3cm},clip]{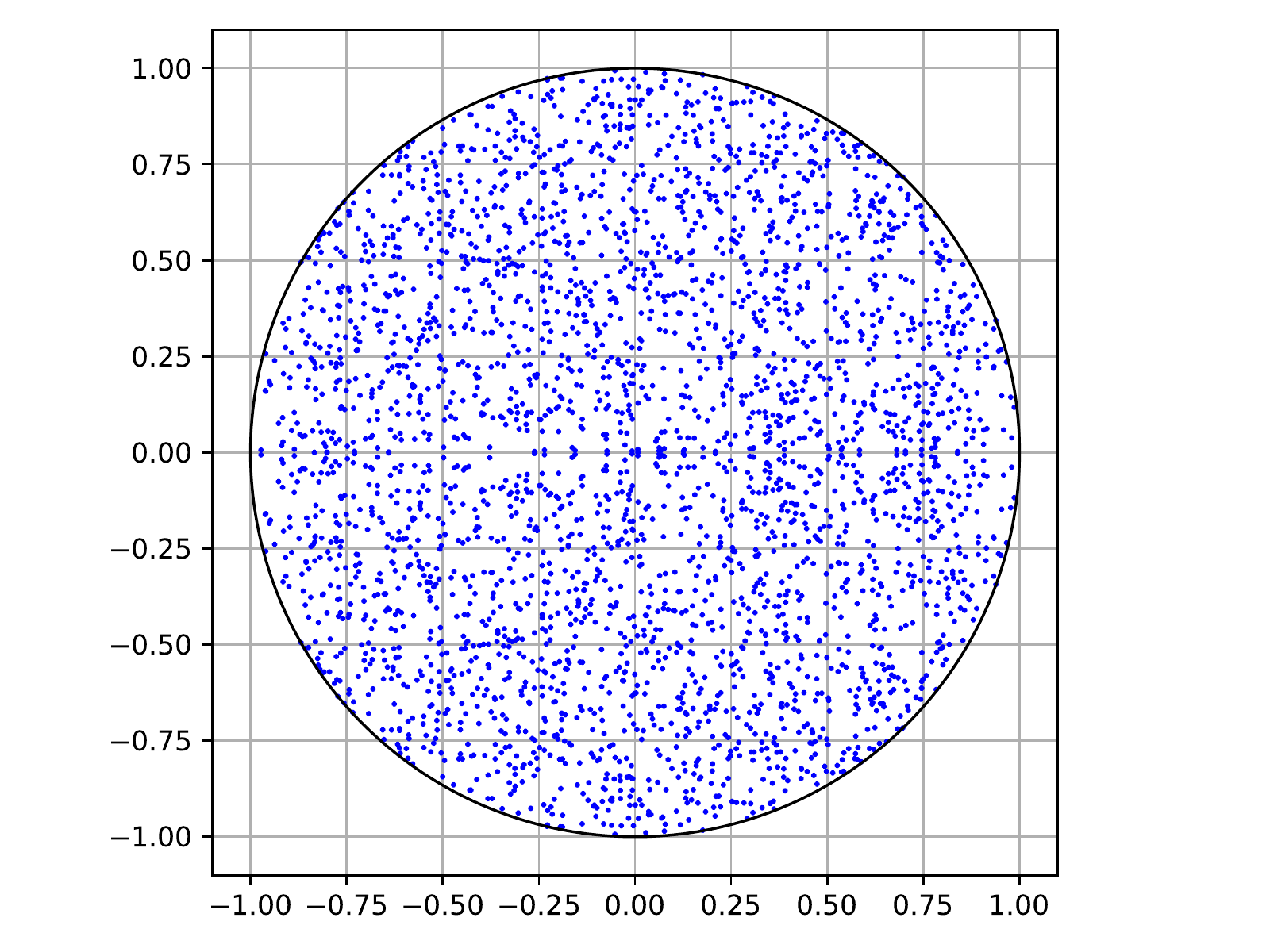} \\ (C)
    \end{minipage}
    \begin{minipage}{.16\linewidth}
        \centering
        \includegraphics[width=\linewidth,trim={1.7cm 0.3cm 2.6cm 0.3cm},clip]{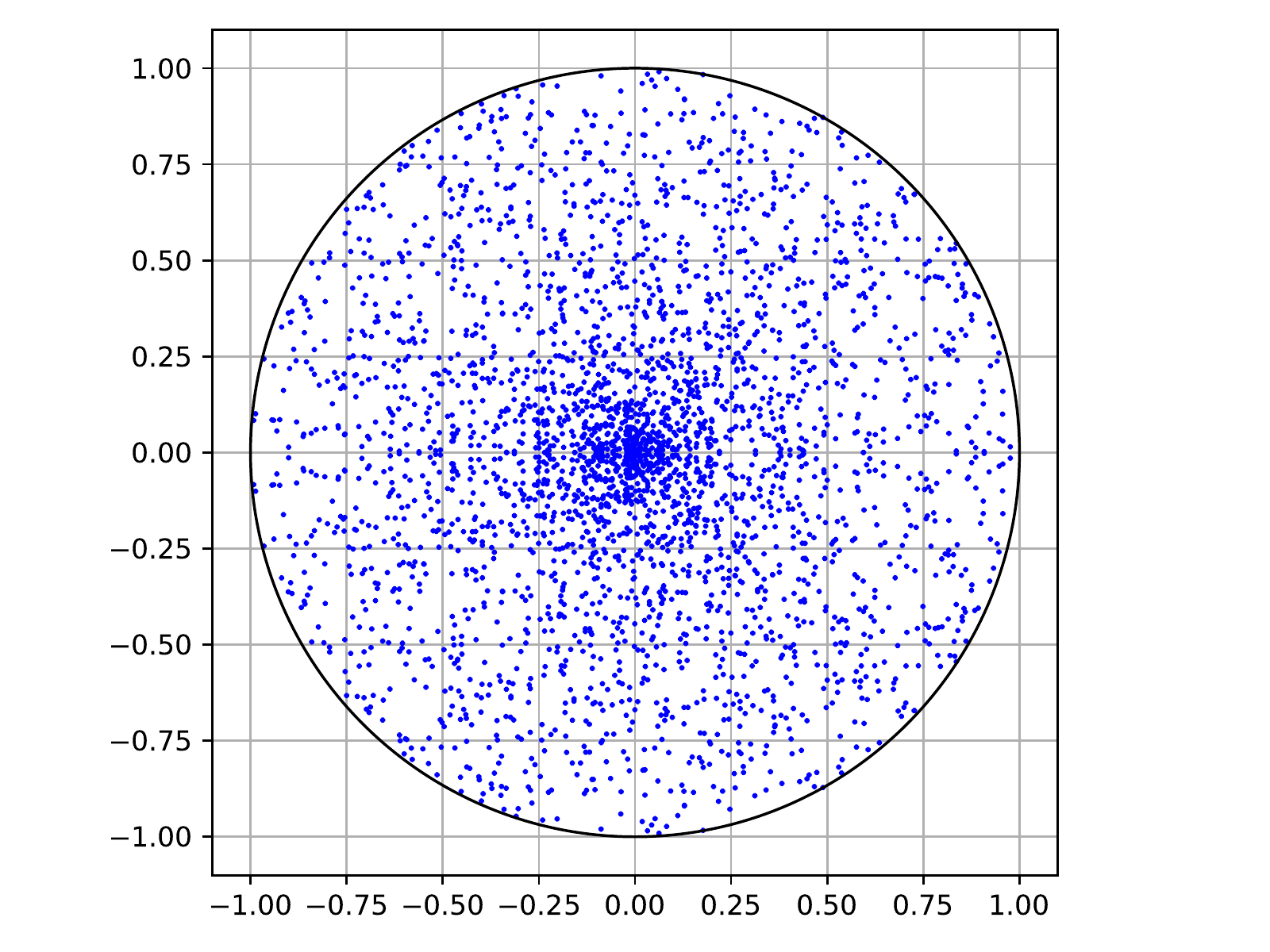} \\ (D)
    \end{minipage}
    \begin{minipage}{.16\linewidth}
        \centering
        \includegraphics[width=\linewidth,trim={1.7cm 0.3cm 2.6cm 0.3cm},clip]{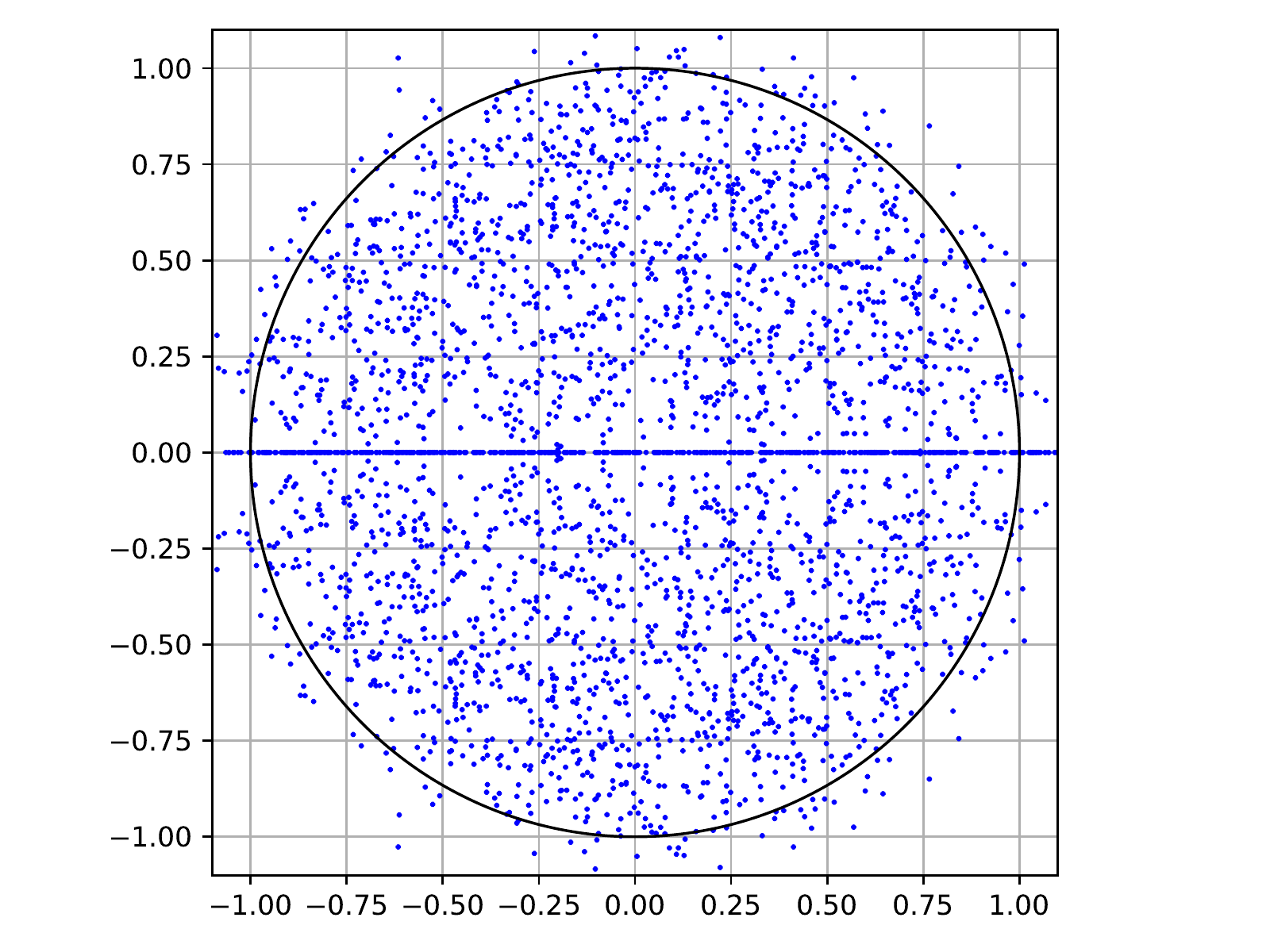} \\ (E)
    \end{minipage}
    \begin{minipage}{.16\linewidth}
        \centering
        \includegraphics[width=\linewidth,trim={1.7cm 0.3cm 2.6cm 0.3cm},clip]{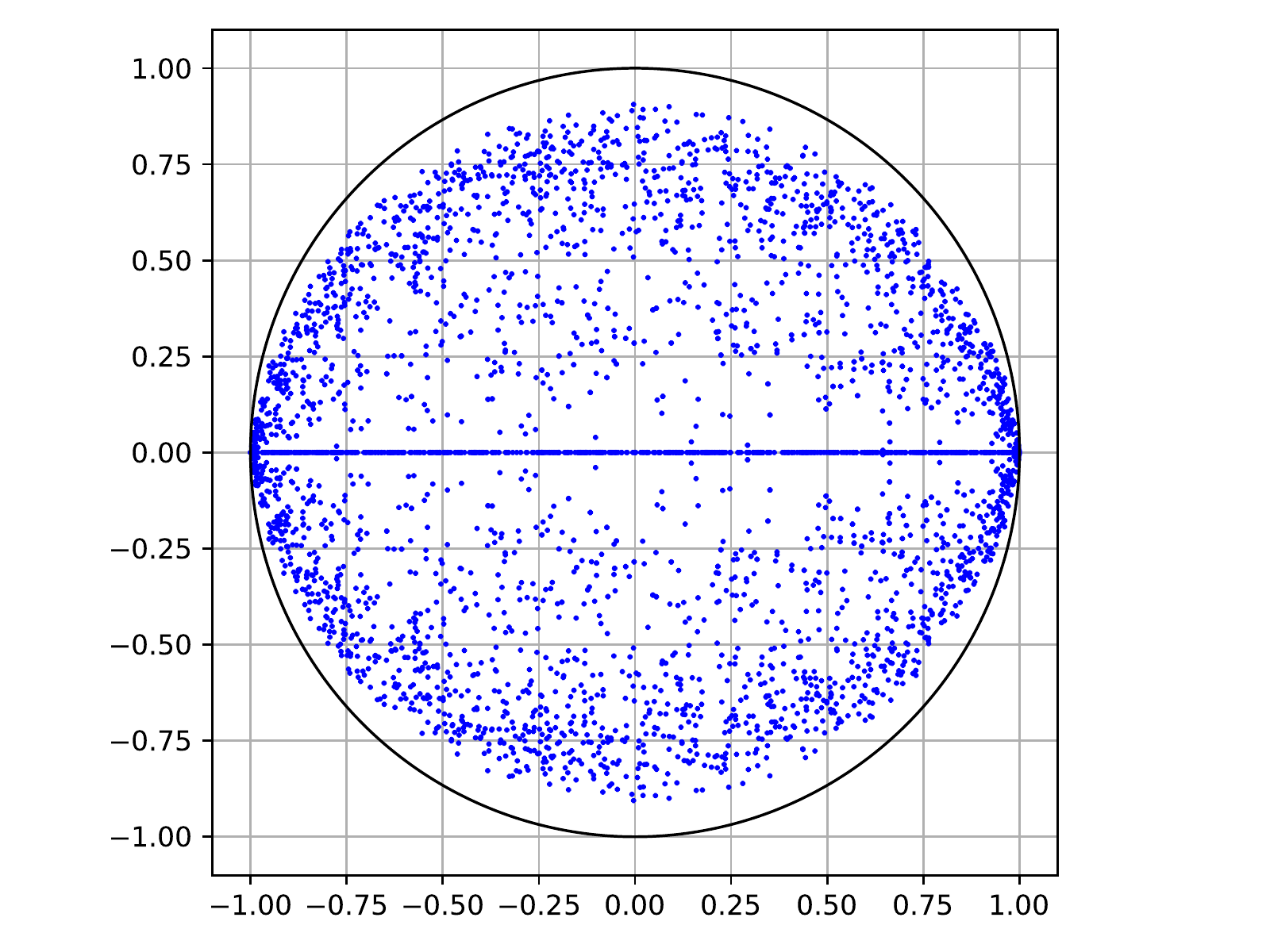} \\ (F)
    \end{minipage}
    \vspace{-0.2cm}
    \caption{Root placements of 100 randomly sampled $32^{\text{nd}}$ degree polynomials. (A) Normal Coefficients, (B) Normal Biquads, (C) Uniform Disk, (D) Uniform Magnitude, (E) Characteristic Polynomial, (F) Uniform Parametric EQs}\label{fig:samplings}
    \vspace{-0.2cm}
\end{figure*}
\section{Background}
\label{sec:background}

\subsection{Digital IIR Filters with Biquad Representations}
\label{ssec:bf}
An $N^{th}$ degree digital IIR filter can be characterized by its transfer function as shown in (\ref{eqn:xfer_func}).
\begin{equation}
  \label{eqn:xfer_func}
    H(z)=
    \displaystyle\frac{b_{0}+b_{1}z^{-1}+\dots+b_{N}z^{-N}}
    {a_{0}+a_{1}z^{-1}+\dots+a_{N}z^{-N}}
\end{equation}
For most applications, $b_{i},a_{i} \in \mathbb{R}$. 
To facilitate numerical stability, these filters are often implemented as a cascade of $K$ second order biquad section
$ H(z) = \prod_{k=0}^{K-1} H_k(z)$
where
\begin{equation}
  \label{eqn:factored_biquad}
    H_{k}(z)=
    \displaystyle g \frac{1+b_{1,k}z^{-1}+b_{2,k}z^{-2}}{1+a_{1,k}z^{-1}+a_{2,k}z^{-2}}.
\end{equation}
Should the poles and zeros of each biquad fall within the unit circle of the complex plane, the digital IIR filter is said to be \textit{minimum phase}. The magnitude response of these filters $|H(e^{i\omega})|$ can be calculated by evaluating $H(z)$ along the unit circle in the complex plane, and taking the magnitude of the result.
\begin{equation}
  \label{eqn:mag_func}
    |H(e^{i\omega})| = 
    \displaystyle \Big|
    \frac{b_{0}+b_{1}e^{-i\omega}+\dots+b_{N}e^{-iN\omega}}{a_{0}+a_{1}e^{-i\omega}+\dots+a_{N}e^{-iN\omega}} \Big| 
\end{equation}
In practice, the logarithm of this magnitude response is of interest,
\begin{equation}
  \label{eqn:logmag_func}
    \log(|H(e^{i\omega})|)=
    \displaystyle G + \log\Big(\prod_{k=0}^{K-1}~\Big|  \frac{1+b_{1,k}e^{-i\omega}+b_{2,k}e^{-2i\omega}}{1+a_{1,k}e^{-i\omega}+a_{2,k}e^{-2i\omega}} \Big|~\Big).
\end{equation}

\subsection{Random Polynomials}
\label{ssec:rp}
Training neural network filter estimators relies on the generation of a dataset of random digital IIR filters. 
While sampling random filters for this training process may appear straightforward, we found the sampling method plays an important role in generalization.
Implicit in the process of random filter generation is a random sampling of polynomials, as demonstrated in  (\ref{eqn:xfer_func}). 
In this section we define several methods for sampling random polynomials of even degree $N$ and comment on their properties. Figure \ref{fig:samplings} shows the root placements of these polynomials sampling methods for 100 filters with degree $32$.

\linesubsec{A. Polynomials with normal coefficients}
Given a degree $N$ polynomial $c_{N}x^{N}+\dots+c_{1}x+c_{0}$, sample each $c_{i}$ from the normal distribution $\mathcal{N}(0,1)$. For sufficiently large $N$, the roots of this polynomial converge to the unit circle~\cite{hammersley}. Most roots are approximately $\frac{1}{N}$ away from the unit circle~\cite{shepp-vanderbei}, and some roots are approximately $\frac{1}{N^{2}}$ away from the unit circle \cite{michelen2020random}. Roughly $\frac{2}{\pi}\log N$ roots fall on the real line \cite{kac}, most of which are close to $1$ and $-1$. The closest real root to the unit circle is approximately $\frac{1}{N}$ away~\cite{michelen2021real}. Much of the behaviour is unchanged if the coefficients are other distributions ~\cite{ibragimov-maslova,sparo-sur,tao-vu-universality}. Under very general conditions, the zeros of these polynomials experience repulsion~\cite{tao-vu-universality}.  
As long as the polynomial and its derivative are not both likely to be small at the same time, the roots repel each other~\cite{tao-vu-universality}.

\linesubsec{B. Biquads with normal coefficients}
\label{ssec:BiqNC}
Given a desired polynomial order $N$, sample $\frac{N}{2}$ second order polynomials $b_{2,i}x^{2}+b_{1,i}x+b_{0,i}$, with $b_{j,i}$ independently sampled from $\mathcal{N}(0,1)$ and multiply them together. This is a process where roots are sampled independently in pairs, which means the roots of the derivative polynomial are uniformly independent in the same way~\cite{pemantle-rivin,kabluchko}. A Monte Carlo simulation with $10^{8}$ iterations suggests that about $64.8\%$ of roots sampled using this method are real. 

\linesubsec{C. Polynomials with uniformly sampled roots in the unit disk}
\label{ssec:UniDisk}
Given a desired order $N$, sample $\frac{N}{2}$ roots in the complex plane using the following procedure: take $\theta$ uniform in $[0,2\pi]$ and $r = \sqrt{U}$ where $U \sim \mathrm{Unif}[0,1]$. Then, select these roots' complex conjugates as the remaining $\frac{N}{2}$ roots. Similar to (B), the roots of the derivative polynomial are uniformly independent in the same way~\cite{pemantle-rivin,kabluchko}, with no expected density of real roots given this sampling. 

\linesubsec{D. Polynomials with roots sampled uniformly in magnitude and argument}
Given a desired polynomial order $N$, sample $\frac{N}{2}$ roots in the complex plane using the following procedure: take $\theta$ uniform in $[0,2\pi]$ and take $r$ uniform in $[0,1]$. Then, select these roots' complex conjugates as the remaining $\frac{N}{2}$ roots. Similar to (C), the roots of the derivative polynomial are uniformly independent in the same way~\cite{pemantle-rivin,kabluchko}. There is no expected density of real roots given this sampling. Compared to the roots sampled in (C), these roots will exhibit a greater density closer to the origin than the unit circle. 

\linesubsec{E. Characteristic polynomial of a random matrix}
Given a desired polynomial order $N$, take a random matrix $A \in \mathbb{R}^{N \times N}$ whose entries are sampled i.i.d. from $\mathcal{N}(0,1)$ and use its eigenvalues (rescaled by $\frac{1}{N}$) as the roots of the desired polynomial~\cite{ginibre,mehta}. These roots exhibit a repulsion from one another \cite{tao-vu-universality}. The eigenvalues converge to the unit disk for various distributions of entries~\cite{tao-vu-circular}. The characteristic polynomial of this random matrix has roughly $\sqrt{{2 N}/{\pi}}$ real roots~\cite{edelman-kostlan-shub}. 
It is known this behaviour persists for a family of random variables whose first four moments match those of the Gaussian (i.e. $E[X] = 0, E[X^2] = 1, E[X^3] = 0, E[X^4] =3$)~\cite{tao-vu-real-evals}, but still open in cases such as if the coefficients are $\{-1,+1\}$ with equal probability~\cite{vu2020recent}.

\linesubsec{F. Uniform parametric EQ}
Given a desired polynomial order $N$, uniformly sample the parameters of a parametric EQ made up of one low shelf section, one high shelf section, and $\frac{N-4}{2}$ peaking filters~\cite{nercessian2020neural}. 
The uniformly sampled parameters include each section's corner/center frequency, gain, and Q factor.

\section{Model Design}
\label{sec:method}

Our goal is to train a neural network to learn a mapping $f_\theta$ that takes a desired magnitude response $X \in \mathbb{R}^F$ sampled at $F$ linearly spaced frequencies over $[0,\frac{f_s}{2}]$, where $f_s$ is the system sample rate, and estimates an $N^\textrm{th}$ order digital IIR filter with a magnitude response $\hat{X} \in \mathbb{R}^F$. 
We fixe $N$ to be even.
This cascade of biquads can then be represented by a scalar gain $G \in \mathbb{R}$, and a set of $K$ second-order sections comprised of $K$~complex poles $P=\{p_{0},...,p_{K-1}~|~p_{k} \in \mathbb{C}\}$ and $K$~complex zeros $Z=\{z_{0},...,z_{K-1}~|~z_{k} \in \mathbb{C}\}$, where $K=N/2$. 
Thus the network learns a mapping $f_{\theta}(X) \rightarrow {G,P,Z} \rightarrow \hat{X}$, with $X,\hat{X} \in \mathbb{R}^{F}$ as shown in Figure~\ref{fig:iirnet-arch}. 
Each pole and zero is paired with its complex conjugate to ensure each biquad has real-valued coefficients.
Thus the $k\textrm{th}$ biquad takes the form
\begin{equation}
  \label{eqn:factored_biquad}
    H_{k}(z)=
 \frac{1-2Re(z_{k})z^{-1}+|z_{k}|^2z^{-2}}{1-2Re(p_{k})z^{-1}+|p_{k}|^2z^{-2}}.
\end{equation}

Estimating a system gain $G$ rather than an individual gain for each second order section reduces the total number of parameters without loss of generality, and was found to aid  stability in training higher order models. 
Additionally, we force the system gain $G \in (0,100)$ to aid training stability by applying the sigmoid function to IIRNet gain estimate and then multiplying by 100.
To ensure a minimum phase filter, the estimated poles $p_{k}$ and zeros $z_{k}$ are rescaled according to \cite{nercessian2021lightweight} as shown in (\ref{eqn:rescaling}).
To further stabilize training, a constant $\epsilon = 10^{-8}$ was added to prevent root placement at the origin or on the unit circle.
\begin{equation}
  \label{eqn:rescaling}
  \begin{aligned}
    p_{k} \leftarrow \frac{ (1-\epsilon) \cdot p_{k} \cdot \tanh( \ |p_{k}| \ )}{ | \ p_{k}+\epsilon \ | }\\
    z_{k} \leftarrow \frac{ (1-\epsilon) \cdot z_{k} \cdot \tanh( \ |z_{k}| \ )}{ | \ z_{k}+\epsilon \ | } 
   \end{aligned}
\end{equation}

During training, the network is tasked with minimizing a loss function $\mathcal{L}$ that measures the distance between the input and estimated magnitude response.
We used the mean squared error of the $\log$ of the magnitude responses of the estimated and target magnitude responses over a set of $F$ linearly spaced frequencies $[0,\frac{f_s}{2}]$.
\begin{equation}
    \mathcal{L} = \frac{1}{F}  \Big|\Big|  \log(\hat{X}) - \log(X) \Big|\Big|^2_{2}
\end{equation}
The complex response of each second order section was calculated by performing the discrete Fourier transform on the numerator and denominator polynomials with zero padding, and dividing the result. 
This allows for parallelized computation, as opposed to the sample-based gradient optimization in previous works~\cite{kuznetsov2020differentiable, pepe2020designing}. 
The base IIRNet architecture is composed of $2$ linear layers with hidden dimension $D$, each followed by layer normalization~\cite{ba2016layer} and LReLU with $\alpha=0.2$.
The final layer has no activation, and projects the hidden dimension to the number of filter parameters, which is a function of the filter order. 
We treat the estimation of complex values as the individual estimation of their real and imaginary components. 
\vspace{-0.2cm}
\subsection{Baselines}
We considered two baselines to benchmark IIRNet against existing methods: the modified Yule-Walker~\cite{chan1982spectral} method and a stochastic gradient descent (SGD) method. 
For the SGD approach we used the same biquad parameterization and loss function as \mbox{IIRNet}, but instead randomly initialized a vector with $G,P,Z$ parameters that are optimized over a number of gradient steps using a learning rate of $5\cdot10^{-4}$.
We then varied the number of gradient steps to observe the impact on run-time as well as accuracy.

\newpage
\section{Experiments}
\label{sec:experiments}
\renewcommand{\arraystretch}{0.7}
\setlength{\tabcolsep}{6.5pt}
\begin{table*}[ht] 
\centering
\vspace{-0.5cm}
\begin{tabular}{l r r r r r r r r r r}
\toprule
\multirow{2}{*}{\textbf{Training Method}}  & \multicolumn{7}{c}{\textbf{Random polynomial families}} & \multicolumn{2}{c}{\textbf{Real-world}} & \multirow{2}{*}{\textbf{Avg}} \\ \cmidrule(lr){2-8} \cmidrule(lr){9-10}
& A &  B &  C & D & E & F & G	& HRTF 	& Gtr. Cab. &  \\ \midrule
Modified Yule-Walker ($N=16$) & 12.84 & 32.46 & 16.67 & 124.23 & 6.80 & 1.40 & 19.73 & \textbf{1.19} & 60.86 & 30.69 \\ \midrule
A. Normal coefficients &  \textbf{4.38}  & 6.80  & 6.22  & 23.11  & 1.42  & 1.11  & 5.07  & 1.35  & 6.73 & 6.24 \\
B. Normal biquads &  13.19  &\textbf{2.70}  & 0.21  & 1.29  & 2.14  & 0.57  & 2.64  & 2.40  & 6.86 & 3.55 \\
C. Uniform disk &  193.81  & 328.79  & \textbf{0.08}  & 1.19  & 8.91  & 50.42  & 83.32  & 263.06  & 1203.40 & 237.00 \\
D. Uniform magnitude disk &  175.81  & 279.54  & 0.09  & \textbf{0.54}  & 11.25  & 61.41  & 76.37  & 250.38  & 1111.05 & 218.49\\
E. Characteristic polynomial &  22.95 & 32.66  & 0.35  & 2.44  & \textbf{0.81}  & 0.72  & 6.81  & 11.02  & 138.99 & 24.08\\
F. Uniform parametric EQ &  19.33  & 12.84  & 3.06   & 17.84  & 3.52  & \textbf{0.21} & 6.89 & 3.79  & 17.50 & 9.44 \\ \midrule
G. All families &  6.24  & 2.89  & 0.11  & 0.67  & 1.12  & 0.34  & \textbf{1.28}  & 1.40  & \textbf{5.59} & \textbf{2.18} \\
\bottomrule
\end{tabular}
\vspace{-0.2cm}
\caption{Average dB MSE for IIRNet trained using different families of random $16^{\textrm{th}}$ order ($K=8$) filters.}
\label{tab:filter_family}
\vspace{-0.4cm}
\end{table*}

\renewcommand{\arraystretch}{0.7}
\setlength{\tabcolsep}{3.5pt}
\begin{table}[ht]
    \begin{tabular}{l r r r r r}
    \toprule
    \multirow{2}{*}{\textbf{Method}}  & \textbf{Params.} & \textbf{Time}& \textbf{(G)} & \textbf{HRTF} & \textbf{Gtr. Cab.}     \\
     & \footnotesize{Million} & \footnotesize{ms} & \footnotesize{dB MSE} & \footnotesize{dB MSE} & \footnotesize{dB MSE} \\ \midrule
        MYW	    & - &   9.00       &     19.73	&    1.19   & 60.86 \\	\midrule
        SGD (1)         & - &    7.75       &  2458.28   &  3165.43   &  5648.83	 \\ 
        SGD (10)        & - &   58.21       &   998.20   &    1393.29   &   2362.49	 \\
        SGD (100)       & - &  578.52       &    11.74   &    3.49   &   5.67	 \\
        SGD (1000)      & - & 5784.94       & 9.49   &    0.76   &   2.25	 \\ \midrule
        IIRNet  64      & 0.04 &    0.28       &     3.70	&    2.74   &    7.22		 \\
        IIRNet 128      & 0.09 &    0.29       &     2.95	&    2.41   &    7.11		 \\
        IIRNet 256      & 0.21 &    0.30       &     2.08	&    2.03   &    6.29		 \\
        IIRNet 512      & 0.55 &    0.36       &     1.51	&    1.69   &    6.54		 \\
        IIRNet 1024     & 1.63 &    0.71       &     1.29	&    1.39   &    5.54		 \\
        IIRNet 2048     & 5.35 &    1.87       &     1.16	&    1.52   &    5.02		 \\
        IIRNet 4096    & 19.1 &    4.65      &     1.11	&    1.38   &    5.86		 \\
    \bottomrule
    \end{tabular}
    \vspace{-0.2cm}
    \caption{Comparison of average dB MSE and runtime in milliseconds fitting $16^{\textrm{th}}$ order filters with other IIR filter design methods.}
    \vspace{-0.5cm}
    \label{tab:hidden_dim}
\end{table}
We used AdamW~\cite{kingma2014adam, loshchilov2017decoupled} and train for $500$ epochs with a batch size of 128, where $1$ epoch is defined as $20,000$ random filters, equating to a total of $10$ million filters. 
The target magnitude response was evaluated over $F=512$ linearly spaced frequencies.
To aid stability, we clip all responses $X \in [-128\,\textrm{dB},\textrm{128}\,\textrm{dB}]$ and then scale responses between $[-1,1]$.
All models were trained with an initial learning rate of $10^{-5}$ unless otherwise noted, and we decayed the learning rate by a factor of $1/10$ at $80$ and $95$\% through training.
We applied gradient clipping where the norm of the gradients exceeded $0.9$.
We conducted a set of three experiments to investigate the behaviour of IIRNet, training a total of 19 models.
We provide code for these experiments along with pre-trained models.\footnote{\scriptsize{{\url{https://github.com/csteinmetz1/IIRNet}}}} 

\linesubsec{Filter family}
To investigate the impact of the random filter sampling method we trained 7 models, training each on a different family of random $16^{\textrm{th}}$ order filters (A-F) as described in Section~\ref{ssec:rp}, with the final model trained using all of the families together (G).
For these experiments, each linear layer had $D=1024$ hidden units, and each model was trained to estimate a $16^{\textrm{th}}$ order biquad cascade. 

\linesubsec{Model size}
The size of the linear layers within IIRNet has a direct impact on the inference time, which is of interest for online and real-time applications. 
We investigated the impact of the model size on the run-time and accuracy by training another 7 models using hidden sizes $D \in 64, 128, ..., 4096$.
For these models we trained using an equal number of random filters from all of the families (G).
All timings were performed on CPU and averaged over a total of 1000 runs, using a machine with an AMD Ryzen Threadripper 2920X.

\linesubsec{Filter order}
IIRNet predicts a fixed order filter given a desired magnitude response, which means that a different model must be trained for different filter order estimations. 
To investigate the performance of our approach as a function of the filter order, we trained another 5 models, varying both the order of the random filters used in training, and the filter order estimated by IIRNet.
These models used $D=2048$ hidden units in each linear layer and were trained again with random filters from all families (G).
Since we found training models that estimate higher order filters ($N \geq 32$) more unstable, we trained all of these models with an initial learning rate of $10^{-6}$.

\vspace{-0.2cm}
\subsection{Datasets}

Three different sets of filters were used to evaluate the models.
First, we evaluated using $1000$ random filters from each of the 7 proposed random filter families (A-G).
We then measured how models generalized to distributions of filters not seen during training, as well as matching the magnitude response of real-world filters such as measured head-related transfer functions (HRTFs) and guitar amplifier cabinets. 
Though phase is an integral part of the HRTF, some studies suggest that the HRTF can be reproduced within perceptual tolerance under certain conditions via a minimum phase magnitude response plus delay match~\cite{kulkarni1999sensitivity}.
Guitar cabinets combine loudspeakers with guitar amplification circuits for use in creative settings within music production. 
The impulse response of these cabinets can then be used for digital emulation of the linear behaviour of these devices. 
In our experiments, 187 HRTFs were sourced from the IRCAM-Listen HRTF Dataset\footnote{\scriptsize{\url{http://recherche.ircam.fr/equipes/salles/listen/}}} and 32 guitar cabinet impulse resposnes were sourced from Kalthallen Cabs\footnote{\scriptsize{\url{https://cabs.kalthallen.de}}}.
All impulse responses were resampled to 16-bit 44.1kHz and a Savitzky-Golay filter~\cite{luo2005savitzky} was used to smooth the magnitude responses before input to IIRNet. 



\section{Results \& Discussion}
We report the MSE between the estimated and target response using a dB scale, enabling a more interpretable analysis of the error.
Experiments with different random filter families in Table~\ref{tab:filter_family} show that training on a specific family of random filters resulted in the best performance when evaluating on that filter family. 
Furthermore, we found that training on certain families (A, B, F) rather than others (C, D, E) resulted in better performance on real-world filters. 
This supports our claim that the method for constructing random filters is a significant consideration in training this type of model.
Notably, IIRNet trained on all filter families (G) achieved the lowest combined MSE across all datasets, indicating that training on multiple families is superior to training on any single family alone.

We also compared performance of these models against the MYW approach, as shown in the first row of Table~\ref{tab:filter_family}. 
Here we used MYW to fit the desired response specifying the filter order $N=16$, the same as the target filter. 
This approach performs worse than IIRNet trained with (G) across all of the random polynomial families, along with the guitar cabinet responses. 
However, we find that MYW outperforms other methods on the HRTF dataset. 
These results point to MYW performing better when the overall range of the magnitude response is more limited, but this approach may struggle when the response has a much larger range in the magnitude space.

The run-time and accuracy of variants of IIRNet are compared to an SGD and MYW approximation on identical datasets in Table \ref{tab:hidden_dim}. 
Both the run-time and accuracy increase as we increase the size of IIRNet, as expected. 
All versions of IIRNet are both faster and more accurate across the set containing all random filter families (G) as compared to both SGD and MYW. 
On the real-world filter estimation tasks, SGD with 1000 iterations outperforms MYW and even the largest IIRNet model, but has a run time orders of magnitude higher.
MYW beats all other approaches on the HRTF estimation task, but performs worse than even the smallest IIRNet model across all random filters families and guitar cabinet estimation. 

Since IIRNet is trained to estimate filters of a fixed order, we evaluated how performance changed as a function of the estimation order.
Table~\ref{tab:filter_ord} demonstrates that in general, increasing the estimated filter order of IIRNet improves estimation accuracy at all orders less than or equal to the training order. 
However, we found training models that estimate filters with order $N\geq64$ challenging, often leading to instability.
As a result, the model trained to estimate $64^{\textrm{th}}$ order filters diverged, and hence performs worse than the $32^{\textrm{nd}}$ order model.


\renewcommand{\arraystretch}{0.7}
\setlength{\tabcolsep}{3.5pt}
\begin{table}[t]
    \begin{tabular}{c r r r r r r r }
    \toprule
    \textbf{Train} & \multicolumn{5}{c}{\textbf{Test order (G)}} & \multirow2{*}{\textbf{HRTF}} & \multirow2{*}{\textbf{Gtr. Cab.}} \\ \cmidrule(lr){2-6}
    \textbf{Order}            & 4 & 8 & 16 & 32 & 64 \\ \midrule
    4  & 1.21 & 7.65 & 20.30 & 75.28 & 196.19 & 11.77 & 19.20 \\
    8  & 0.37 & 1.59 & 6.20 & 24.98 & 80.10 & 6.08 & 11.96\\
    16 & 0.22 & 0.68 & 2.13 & 9.55 & 34.76 & 1.97 & 6.12\\
    32 & \textbf{0.17} & \textbf{0.39} & \textbf{0.98} & \textbf{4.82} & \textbf{21.32} & \textbf{0.66} & 
    \textbf{1.92}\\
    64 & 1.96 & 2.07 & 2.69 & 7.49 & 22.61 & 3.29 & 4.70\\
    \bottomrule
    \end{tabular}
    \vspace{-0.2cm}
    \caption{dB MSE evaluating IIRNet with different estimation orders.} 
    \label{tab:filter_ord}
    \vspace{-0.5cm}
\end{table}

While our results demonstrate that IIRNet produces accurate estimates of both unseen random and real-world filters, this approach has some limitations including fixed order filter estimates, consideration only of the magnitude response, and the inability to apply additional design constraints.
Future work could investigate a formulation of the loss function that also considers phase, along with architectural adjustments that may support variable filter order estimation.
However, it may be possible to address some of these limitations by using IIRNet simply as a method for generating an initial estimate that can be refined with more flexible iterative techniques. 

\vspace{-0.2cm}
\section{Conclusion}
\label{sec:conclusion}
\vspace{-0.2cm}
We presented IIRNet, a neural network for the direct estimation of the biquad filter cascades to match an arbitrary magnitude response.
We investigated performance of IIRNet using a diverse range of random filter families informed by knowledge of random polynomials. 
Performance was measured using a large dataset of random filters, head-related magnitude responses, and guitar cabinet magnitude responses. 
We demonstrated that using a variety of random sampling methods performs best across datasets, outperforming all models trained with only a single random filter family. 
IIRNet is shown, on average, to perform faster and more accurate filter estimation compared to modified Yule-Walker and stochastic gradient descent, requiring no manual parameter tuning during the design process. 
Additionally, we demonstrate the accuracy-speed trade-off when varying the network size, and show how training with higher order filters produces superior generalization performance across tasks.


\section{Acknowledgement}\vspace{-0.2cm}
JC is supported by the Research and Development Division of Yamaha Corporation, Japan. CS is supported by EPSRC UKRI CDT in Artificial Intelligence and Music (EP/S022694/1). MM is supported in part by NSF grant DMS-2137623.
\vspace{-0.1cm}

\bibliographystyle{IEEEbib}
\bibliography{strings,refs}

\end{document}